# DOES THE PRINCIPLE OF EQUIVALENCE PROHIBIT TRAPPED SURFACES FROM FORMING IN THE GENERAL RELATIVISTIC COLLAPSE PROCESS ?

Darryl Leiter (1), Stanley Robertson (2)

It has been recently shown [1] that time-like spherical collapse of a physical fluid in General Relativity does not permit formation of ``trapped surfaces''. This result followed from the fact that the formation of a trapped surface in a physical fluid would cause the time-like world lines of the collapsing fluid to become null at the would-be trapped surface, thus violating the Principle of Equivalence in General Theory of Relativity (GTR). For the case of the spherical collapse of a physical fluid, the "*no trapped surface condition*" $2GM(r,t) / R(r,t)c^2 < 1$ was found to be required to be satisfied in all regions of space time, where R(r,t) is the invariant circumference variable, r is a co-moving radial coordinate and M(r,t) is the gravitational mass confined within the radius r. The above result was obtained by treating the problem from the viewpoint of an internal co-moving observer at radius r. The boundary of the fluid at $r_S= R_S(r_S , t)$ must also behave in a similar manner, and an external stationary observer should be able to obtain a similar "no trapped surface" relationship. Accordingly, we generalize this analysis by studying the problem of a time-like collapsing radiating plasma from the point of view of the *exterior stationary observer.* We find that the Principle of Equivalence implies that the physical surface surrounding the plasma must obey $1 / (1+ z_S) > 0$ where $z_S$ is the surface red shift. When this condition is applied to the first integral of the time-time component of the Einstein Equation it leads to the "no trapped surface condition" $2GM(r_S,t) / R(r_S,t)c^2 < 1$ consistent with the condition obtained above for the interior co-moving metric. The Principle of Equivalence enforces the "*no trapped surface condition*" by constraining the physics of the general relativistic radiation transfer process in a manner which requires it to establish and maintain an Eddington limited secular equilibrium on the dynamics of the collapsing radiating surface so as to always keep the physical surface of the collapsing object outside of its Schwarzschild radius. The important physical implication of the "*no trapped surface condition*" is that galactic black hole candidates GBHC *do not* possess event horizons and hence *do* possess intrinsic magnetic fields. In this context the spectral characteristics of galactic black hole candidates offer strong evidence [2] that their central nuclei are highly red shifted Magnetospheric Eternally Collapsing Objects (MECO), within the framework of General Relativity.

KEYWORDS: Black Holes, Event Horizons, Trapped Surfaces, General Relativity

(1) FSTC, Charlottesville, Virginia. 22901, USA

(2) Physics Department, Southwestern Oklahoma State University, Weatherford, O.K.

**I. INTRODUCTION**

Recent observations of the quiescent luminosities of Neutron Stars (NS) and Galactic Black Hole candidates (GBHC) have suggested that both NS and GBHC possess intrinsic magnetic moments [2]. Accommodating intrinsic magnetic moments in such objects will clearly require revisions in the current theoretical description of these compact objects in the context of General Relativity because if GBHC are Black Holes with Event Horizons they could not exhibit the effects of any internal magnetic field. Albert Einstein attempted to show that Black Holes with Event Horizons were not allowed by the General Theory of Relativity [3] , however his argument lacked the complete theoretical formalism associated with the concept of black holes and gravitational collapse. Using the currently known general formalism for spherical collapse in the context of an interior co-moving observer, and noting that the Principle of Equivalence requires that the worldline of timelike matter trajectories must remain time-like in all regions of space time [4], it has been recently shown [1] that no trapped surfaces can be formed in spherical gravitational collapse. While the above result was obtained by treating the problem from the viewpoint of an internal co-moving observer at radius r, the boundary of the fluid at $r_S = R_S(r_S , t)$ must also behave in a similar manner, and an external stationary observer should also observe the same no trapped surface relationship.

Since the Principle of Equivalence does not allow the time-like world line of physical matter, being acted on by both gravitational and non-gravitational forces, to become null in any regular region of space time, we show in section II that this requires that the condition $1 / (1+z) > 0$ must hold, where z is the red shift of the physical matter moving along its world line as seen by a distant observer. In section III adopt the viewpoint of a stationary observer in the exterior metric of a time-like collapsing plasma which is radiating photons while obeying the Principle of Equivalence requirement that $1 / (1+z_S) > 0$ must always hold. Then the Einstein equation can be solved with an appropriate energy momentum tensor and boundary conditions for the metric. When the Principle of Equivalence requirement that $1 / (1+z_S) > 0$ is applied to the first integral of the mixed time-time component of the Einstein equation, with appropriate boundary conditions on the time-like surface surrounding the collapsing radiating plasma, we find that it requires that the "no trapped surface condition" $2GM_S / R_S c^2 < 1$ must hold on the physical boundary of the collapsing radiating plasma , consistent with earlier work done for so-called Eternally Collapsing Objects (ECO) in the context of a co-moving observer in an interior metric[1]. In Appendix A we discuss this point in more detail and show that in the context of the Einstein equations for a collapsing, radiating plasma, the Principle of Equivalence (POE) requires that the physics associated with the general relativistic radiation transfer process in the MECO must dynamically establish an Eddington limited secular equilibrium condition on the collapsing surface of the MECO before the collapsing surface has a chance to pass thru the Schwarzschild radius of the collapsing object. Finally we consider the observational implications of this result for intrinsic magnetic moments for GBHC [2] and argue that the non-existence of trapped surfaces becomes manifest observationally in the fact that the spectral properties of the radiation emitted by GBHC can be consistently understood in terms of a unified model of Magnetospheric Eternally Collapsing Objects (MECO) within the framework of General Relativity.

## II. THE PRINCIPLE OF EQUIVALENCE IN GENERAL RELATIVITY REQUIRES THAT PHYSICAL MATTER OBEYS TIME-LIKE WORLDLINES IN SPACETIME

In General Relativity the Strong Principle of Equivalence requires that Special Relativity must hold locally for all freely falling time-like observers. Hence the Principle of Equivalence requires that the proper time $d\tau > 0$ along time-like space time world lines of physical matter with non-zero rest mass must always be positive definite so that the physical motion of matter will always remain time-like in all of spacetime. Since this requirement can be expressed as a tensor relationship this means that in a general curved space time physical matter must follow time-like world lines such that the associated invariant line interval respectively obeys

$$ds^2 = g_{\mu\nu}\, dx^\mu\, dx^\nu = c^2 dt^2 (g_{\mu\nu}\, v^\mu v^\nu / c^2) = c^2 d\tau^2 > 0 \qquad (1)$$

where $v^\mu = dx^\mu / dt$ is the *coordinate 4- velocity* , $d\tau$ is the infinitesimal proper time. For the case of physical matter, the above condition (1) can be written in a more physically transparent form [4] as

$$d\, s^2 = c^2 d\tau^2 = [c^2 d\tau_{syn}{}^2 - dl^2] > 0 \qquad (2\text{-a})$$

where $d\tau_{syn}$ is the element of proper time synchronized along the particle trajectory given by

$$d\tau_{syn} = [\, |1 + (g_{ok}/g_{oo})\, (v^k / c)|\, ]\, (g_{oo})^{1/2}\, dt \qquad (2\text{-b})$$

and $dl$ is the element of proper length given by

$$dl = [(g_{ok}\, g_{oj} / g_{oo} - g_{kj})\, dx^k\, dx^j\, ]^{1/2} = [(\gamma_{kj})\, dx^k\, dx^j\, ]^{1/2} \qquad (2\text{-c})$$

Since the *physical 3-velocity* $V^k$ can be defined in terms of the proper time $d\tau_s$ as

$$V^k = dx^k / d\tau_{syn} = v^k / [|1 + (g_{oi} / g_{oo}) (v^i / c)|] (g_{oo})^{1/2} \quad \text{(2-d)}$$

Then the *physical speed* , defined in terms of the proper length dl as $V = (dl / d\tau_{syn})$ , is

$$V = [(g_{ok} g_{oj} / g_{oo} - g_{kj}) v^k v^j]^{1/2} / [[|1 + (g_{oi} / g_{oo}) (v^i / c)|](g_{oo})^{1/2}] = [(\gamma_{kj}) V^k V^j]^{1/2} \quad \text{(2-e)}$$

and from (2-e) and (2-a)

$$ds^2 = c^2 d\tau^2 = c^2 d\tau_{syn}{}^2 [1 - V^2 / c^2 ] > 0 \quad \text{(2-f)}$$

Hence $d\tau$ and $d\tau_{syn}$ are related by

$$d\tau = d\tau_{syn} [1 - V^2 / c^2]^{1/2} > 0 \quad \text{(2-g)}$$

Since the *proper 4-velocity* is $u^\mu = dx^\mu / d\tau$ , then the *proper 3-velocity* $u^k = dx^k / d\tau$

is related to the *physical 3-velocity* $V^k = dx^k / d\tau_{syn}$ by

$$u^k = [V^k ] / [1 - \{V / c\}^2]^{1/2} \quad \text{(2-h)}$$

while time component of the proper velocity $u^o = dx^o / d\tau$ is given by

$$u^o = [c (g_{oo})^{-1/2} + (g_{ok} / g_{oo}) V^k ] / [1 - \{V / c\}^2]^{1/2} \quad \text{(2-i)}$$

Recalling from (2-e) that the *physical speed* is defined as $V = [(\gamma_{kj}) V^k V^J]^{1/2}$ then in an

analogous manner we define the *proper speed* as $u = [(\gamma_{kj}) u^k u^k]^{1/2}$ and find that

$$[1 - \{V / c\}^2] = 1 / [1 + \{u / c\}^2] > \quad \text{(2-j)}$$

Hence from (2-f) or (2-g) we find that for time-like matter trajectories, associated with matter having non-zero physical speed V, the proper time is given by

$$d\tau = (ds / c) = dt \{(g_{oo})^{1/2} [|1 + (g_{ok}/g_{oo}) (v^k / c)|] (1 - V^2/ c^2 )^{1/2}\} > 0 \quad (2\text{-k})$$

$$= dt \{(g_{oo})^{1/2} [ |1 + (g_{ok}/g_{oo}) (v^k / c)| ] (1 + u^2 /c^2)^{-1/2}\} > 0 \quad (2\text{-j})$$

$$= dt \{1 / (1 + z)\} > 0 \quad \text{where z is the red-shift associated with time dilation} \quad (2\text{-l})$$

The above equations are valid even in the case where the metric is time dependent, and in the special case where the velocities vanish reduce to the well-known result that $d\tau = dt\, (g_{oo})^{1/2}$

Hence we see from equations (2) that the Principle of Equivalence(POE) requirement that the proper time $d\tau > 0$ along time-like world lines of physical matter in space time must exist and always remain time-like implies that:

$$1 / (1 + z) > 0 \text{ , where z is the red-shift associated with time dilation}$$

$$\text{or equivalently that both} \quad g_{oo} > 0 \quad \text{and} \quad 0 \leq V/ c < 1 \quad (3)$$

must hold for along all time-like world lines of physical matter over all regions of spacetime[1].

This POE condition requires that the non-geodesic timelike matter trajectories derived from $T^{\mu\nu}{}_{;\nu} = 0$ must have the property that they can be extended to any value their affine parameter. A spacetime manifold which has this property is called “bundle complete” [5]. In order to insure that this property holds in the resulting spacetime continuum, the Principle of Equivalence acts as a constraint on the physical structure of the energy momentum transport described by the energy momentum tensor $T^{\mu\nu}$ on the right hand side of the Einstein equation $G^{\mu\nu} = (8\pi G / c^4)\, T^{\mu\nu}$. In the following sections we will find that this requires that the energy momentum tensor, which describes time-like collapsing compact objects containing mass $M_S$ and surrounded by a physical surface of radius $R_S$, must always contain components which allow it radiate photons at the Eddington limit.

## III. THE PRINCIPLE OF EQUIVALENCE PREVENTS TRAPPED SURFACES FROM BEING FORMED BY REQUIRING THAT GRAVITATIONALLY COLLAPSING PLASMAS MUST RADIATE PHOTONS AT THE EDDINGTON LIMIT

We have shown that a strict application of the Principle of Equivalence to the solutions of the Einstein field equations in General Relativity, implies that event horizons cannot be physically realized through the time-like collapse of radiating physical matter. The argument was based on the fact that if an event horizon were to form during collapse, this would require the time-like worldline of the collapsing matter to become null in violation of the Principle of Equivalence in General Relativity**.** Although the existence of objects compact enough to qualify as black hole candidates is beyond question it is still observationally unclear whether event horizons can be physically realized in the collapse of stellar mass objects. No clear-cut observational evidence for the existence of event horizons has yet been reported. On the other hand, if event horizons are not formed, these collapsed objects should retain substantial interior magnetic moments which should generate large magnetic fields in their exterior. Hence observations of the effects these magnetic fields on accreting matter in the exterior regions could resolve the issue of whether event horizons exist.

Now in analogy to the case of the Vaidya metric associated with pure neutrino emission studied earlier [6,7,8,9] , the radiating metric associated with this more general scenario involving photons will have the general form

$$ds^2 = A^2 (r,u) (c^2dt^2) + 2 D(r,u) c\, dt\, dR - B^2 (r,u) dR^2 - R(u)^2[d\theta^2 + \sin\theta^2 d\phi^2] \quad (4\text{-a})$$

where $u = t - R/c$ is the retarded time. Substitution of the metric (4-a) into the Einstein equation

$$G\mu^{\nu} = (8\pi G / c^4)\, T\mu^{\nu} \qquad (4\text{-}b)$$

with an energy momentum tensor $T\mu^{\nu}$ whose form describes the radiating plasma collapse scenario leads to a set of equations from which, with appropriately chosen boundary conditions, the functions A(r,u), B(r,u), and D(r,u ) can be determined. However independent of the detailed form of the $T\mu^{\nu}$, the first integral of the mixed time-time component of the Einstein equation $Go^{o} = (8\pi G / c^4)\, To^{o}$ , with appropriately chosen boundary conditions on the time like surface S separating the exterior metric from the interior metric, leads to the proper time of the collapsing surface S synchronized along world line of the collapsing plasma on the surface of invariant radius $R_S$

$$d\tau_S = du \;/\; (1 + z_S) = du\,(\Gamma_S + U_S / c) > 0 \qquad (5\text{-}a)$$

where z the red shift of the radiating collapsing surface, $u = t - R_s / c$ and

$$\Gamma_S = [1 - R_{Schw} / R_S + (U_S / c)^2]^{1/2} \qquad (5\text{-}b)$$

where $U_S = (dR / d\tau)_S < 0$ is the proper time rate of change of the invariant radius $R_S$ and $R_{Schw} = 2GM_S / c^2$ where $M_S(r, t)$ is the total gravitational mass contained within the surface of radius $r = R_S$ is defined by the general formula

$$M_S (R_S , t) = \int( T_o^{o}\, \Gamma\, d\mathbf{V} )_S \qquad (5\text{-}c)$$

where $\Gamma_S = (dR / dl)_S$ and $d\mathbf{V}$ is the proper volume element

$$d\mathbf{V} = 4\pi R^2 dl \qquad (5\text{-d})$$

and dl is the proper length inside and on the surface $r = R_S$ is given by

$$dl = (B\, dr) \qquad (5\text{-e})$$

Assuming that the $M_S$ given by (5-c) is being radiated away by photon emission whose luminosity at infinity is given by

$$L(\infty) = - dM_S c^2 / du = - [ dM_S c^2 / d\tau ] (1 + z_S) \qquad (6\text{-a})$$

then the proper time rate of change of the Schwarzschild radius $R_{Schw} = 2GM_S / c^2$ associated with the mass $M_S$ inside of S, is

$$U_{Schw} = (dR_{Schw} / d\tau)_S = (2G/ c^4) dM_S c^2 / d\tau_S < 0 \qquad (6\text{-c})$$

Now since the Principle of Equivalence requires that the collapse of the radiating physical surface S must always be time-like, this requires that the element of proper time along the radial world line of a fluid element on the boundary of the collapsing fluid must always obey $d\tau_S > 0$. Hence the Principle of Equivalence requires that

$$d\tau_S = du ([1 - R_{Schw} / R_S + (U_S / c)^2]^{1/2} + U_S / c) = du / (1 + z_S) > 0 \qquad (7)$$

which for $U_S < 0$ requires that the "no trapped surface condition" $R_{Schw} / R_S < 1$ must hold.

In order for the Principle of Equivalence requirement (7) to be satisfied by a collapsing red-shifted spherical surface of radius $R_S > R_{Schw}$ containing plasma with gravitational mass $M_S$ , the physics associated with the radiation transfer within the collapsing plasma must eventually allow it to start emitting photons at the local Eddington limit $L_{Edd}$ . This is because under these conditions the local radial proper time rate of change of the radius of the surface ($U_S$ / c) can be vanishingly small and the Principle of Equivalence "no trapped surface condition" requirement $R_S > R_{Schw}$ for the collapse can be dynamically satisfied by the equations of motion of the matter.

Hence the Principle of Equivalence requires that the physical dynamics associated with the general relativistic radiation transfer process must establish an Eddington limited secular equilibrium on the collapsing surface(2) (3) before it has a chance to pass thru the Schwarzschild radius of the collapsing object. In Appendix A we show in detail that when the Eddington limit is established at red shift (1+z) = (1+ $z_{Edd}$) , the Principle of Equivalence applied to the Einstein Equation implies that the time like collapsing radiating surface of the MECO lies outside of the Schwarzschild radius of the collapsing object and remains that way for the duration of the Eddington limited collapse process. In Appendix B we show that if $L_{Edd}(\text{escape})_S$ is the luminosity generated by the interior physics of the which escapes through the collapsing MECO surface, the red shift at which an Eddington balance will be achieved is given by

$$(1 + z_{Edd}) = [L_{Edd}(\text{escape})_S / [4\pi G M_S c / \kappa ] \qquad (8\text{-a})$$

Then terms of the quantity $U_{Sch} = (dR_{Sch} / d\tau)$ this implies that

$$(U_{Schw}/c) = - (2G / c^5) [L_{Edd}(escape_S)/ (1 + z_{Edd})] < (U_S/c) = 0 \qquad (8\text{-b})$$

consistent with the Principle of Equivalence requirement that $R_{Schw} < R_S$ must be maintained by the physical forces which act during the collapse process.

The dynamic radiation transfer condition required for the existence of the high redshift Eddington limited stationary equilibrium, associated with the MECO gravitational plasma collapse process satisfying the Principle of Equivalence, is that the MECO must contain an intrinsic equipartition magnetic field energy density which is less than or equal to its gravitational field energy density. Since both the local proper energy and local proper volume are proportional to $(-g_{rr})^{1/2}$, the stationary equilibrium condition for the MECO is $(1/8\pi) B^2_{Equip} \leq (1/4\pi G) (GM_S /R_S{}^2)^2$. This requires the MECO to contain an intrinsic equipartition magnetic dipole field given by

$$B_{EQUIP} = (c^2/ (20G)^{1/2}) (R_{Schw} / R_S{}^2)$$

$$= (6 \times 10^{18} \text{ Gauss}) / M^* \qquad (9)$$

where $M^* = M / Mo$ and Mo is the solar mass in gm.

Hence it is this intrinsic equipartition magnetic dipole field $B_{EQUIP}$ in the MECO which generates the required out flowing Eddington limited luminosity $L_{Edd}(outflow)_S$ at the surface S in the form of optically thick synchrotron radiation. Under these conditions stability occurs since a decrease (increase) of $(U_S / c)$ from its Eddington limited value this causes a corresponding decrease (increase) in $R_S$ which in turn causes an increase (decrease) in $B_{EQUIP}$ and $L_{Edd}(outflow)_S$ which counteracts the change in $(U_S / c)$ and brings it back to its Eddington limited value. The strength of the intrinsically equipartition magnetic fields $B_{EQUIP} \sim (6 \times 10^{18}$ gauss$) / M^*$ at the surface of the MECO(4),(5), when distantly observed are reduced by a factor of $(1 + z_S)$ from their values at the collapsing surface $R_S$ .

In previous work [2] on galactic black hole candidates (GBHC) containing MECO with $M^* = 10$, the distantly observed magnetic fields were $\sim 10^{10}$ gauss which implied that $(1 + z_S) \sim 10^8$. This value of $(1 + z_S) \sim 10^8$ is also consistent with the case of AGN with $M^* = 10^8$ since then the distantly observed magnetic fields would be $\sim 10^4$ gauss consistent with what is expected in the AGN accretion disk environment. Now from Appendix B it can be shown that the net Eddington luminosity escaping from the surface from the MECO $L_{Edd}(escape)$ is given by

$$L_{Edd}(escape)_S = (4\pi G M_S(\tau) c / \kappa) (1 + z_S) \quad \text{erg/sec} \qquad (10)$$

where $(1 + z_S) = 10^8$ . Hence the MECO will have a very high value of the compactness parameter L/R which implies that photon-photon collisions leading to pair production will be dominant in the MECO plasma. This implies that $L_{Edd}(escape)_S$ will be emitted by optically thick synchrotron radiation from an electron-positron dominated plasma, which is proportional to the 4th power of the intrinsic equipartition magnetic field [11] ,where the high compactness of the pair dominated plasma in the MECO effectively acts as a thermostat which buffers the temperature of the optically thick synchrotron radiation escaping from the MECO surface to be $T_S \sim 6 \times 10^9$ K. Because of its high compactness the Eddington limited MECO surface will also be surrounded by an *optically thick radiation dominated pair atmosphere* which, after reprocessing the photons, will allow reprocessed luminosity $L_P(\text{escape})$ to escape from an associated photosphere at redshift $z_P > z_S$ . If $T_P$ is the temperature of the photosphere at redshift $z_P$ , then the reprocessed luminosity $L_P(\infty)$ escaping to infinity from the radiation dominated pair atmosphere inside of the photon orbit is given by

$$L_P(\infty) = L_P(\text{escape}) / (1 + z_P)^2 = 4\pi R_p^2 (\sigma T_P^4) [(27/4) / (1 + z_P)^4] \text{ erg/sec} \quad (11\text{-a})$$

However energy conservation implies that $L_P(\infty) = L_S(\infty)$ where

$$L_S(\infty) = L(\text{escape})_S / (1 + z_S)^2 = (4\pi G M_S(\tau) c / \kappa) / (1 + z_S) \quad \text{erg/sec} \quad (11\text{-b})$$

Now since the radiation dominated pair atmosphere exists in the metric exterior to the MECO surface, integration of the associated stationary Euler equation in hydrostatic balance within the pair atmosphere where $z_S < z < z_P$ implies that the quantity $[P / (1+ z)]$ is approximately constant, where for radition pressure $P = (\sigma T^4) / c$ applies . Hence it follows that the temperature $T_S$ at the bottom and the temperature $T_P$ at the top of the pair atmosphere obey

$$[T_S^4 / (1+ z_S)] = [T_P^4 / (1+ z_P)] \quad (11\text{-c})$$

Solving equation (11-a,b,c) for $T_P$ and $z_P$ . and using the fact that $(1 + z_S) = 10^8$ and $T_S \sim 6 \times 10^9$ K we find that redshift $z_P$ and the temperature $(T_P)_\infty = (T_P) / (1+ z_P)$ of the pair atmosphere's photosphere seen at infinity are given respectively by

$$(T_P)_\infty = (2.35 \times 10^5 \text{ K}) / M^{*1/4} \qquad (12\text{-a})$$

$$(1 + z_P) = (1.63 \times 10^3) M^{*1/3} \qquad (12\text{-b})$$

where in the above we have used

$\kappa = 0.4$ cm$^2$/ gm (Thompson plasma opacity)

$G = 6.67 \times 10^{-8}$ cm$^3$/ gm-sec$^2$ (Gravitational Constant)

$M_S = (2 \times 10^{33}) M^*$ gm (MECO mass in solar mass units)

$\sigma = 5.67 \times 10^{-5}$ erg / sec-cm$^2$-K$^4$ (Stephan-Boltzmann Constant)

For a quiescent 10 solar mass (M* = 10) MECO the surrounding pair photosphere is red shifted by $(1 + z_P) \sim 3.51 \times 10^3$ and the corresponding to a quiescent thermal emission seen at infinity is

$$(L_P)_\infty \sim 1.3 \times 10^{31} \text{ erg/sec}$$

$$(T_P)_\infty \sim 1.32 \times 10^5 \text{ K}$$

$$(\lambda_{Edd})_\infty \sim 220 \text{ Angstroms (hard UV)}$$

For a quiescent $10^8$ solar mass ($M^* = 10^8$) MECO the surrounding pair photosphere is red shifted by $(1 + z_P) \sim 7.56 \times 10^5$ and the corresponding to a quiescent thermal emission seen at infinity is

$$(L_P)_\infty \sim 1.3 \times 10^{38} \text{ erg/sec}$$

$$(T_P)_\infty \sim 2.35 \times 10^3 \text{ K}$$

$$(\lambda_{Edd})_\infty \sim 123 \text{ microns (infrared)}$$

Hence even though they are not black holes quiescent MECO have lifetimes much greater than a Hubble time (see Appendix A and B) while emitting highly red shifted quiescent thermal spectra that may not easily observable. Recently a GBHC model based on an active MECO with an accretion disk [2] has been used to calculate the intrinsic magnetic moments and rates of spin for the GBHC. In this context the important physical result of the "*no trapped surface condition*" associated with their central MECO is that galactic black hole candidates GBHC *do not* possess event horizons and hence *do* possess intrinsic magnetic fields. Hence it follows that the spectral characteristics of galactic black hole candidates offer strong evidence that their central nuclei are highly red shifted Magnetospheric Eternally Collapsing Objects (MECO), within the framework of General Relativity. It was found that this model could be accurately used to predict the observed GBHC quiescent luminosities. The accuracy of these calculations offered strong empirical support for the conclusion, based on the Principle of Equivalence, that trapped surfaces leading to event horizons cannot be formed by the time like collapse process in the GBHC.

## IV. SUMMARY AND CONCLUSIONS

In General Relativity the Strong Principle of Equivalence requires that Special Relativity must hold locally for freely falling time-like observers in all of spacetime. Since this requirement is a tensor relationship it implies that the proper time along the time-like spacetime world lines of physical matter with non-zero rest mass must always be positive definite $d\tau > 0$. In order to satisfy this Principle of Equivalence requirement, the time-like world lines of physical matter under the influence of both gravitational and non-gravitational forces must not become null in any regular region of space time. Spacetime manifolds which have this property are called "bundle complete" ( see pages 69-71 in Wheeler, A., Ciufolini, I., "Gravitation and Inertia" , (1995), Princeton University Press Princeton, New Jersey). Since this requires that the associated time-like matter world lines derived from the Bianchi Identity $T^{\mu\nu}{}_{;\nu} = 0$ must have the property that they can be extended to any value their affine parameter, this Principle of Equivalence requirement dynamically affects the nature of the physically acceptable forms which the energy momentum tensor on the right hand side of the Einstein equations are allowed to take. In this context we found the Strong Principle of Equivalence required that for all time-like world lines of physical matter over all regions of spacetime the following condition must hold:

$1 / (1 + z) > 0$ , where z is the red-shift associated with time dilation

or equivalently that both $g_{oo} > 0$ and $0 \leq V/c < 1$

In order to dynamically insure that this property holds in the resulting spacetime continuum, the Principle of Equivalence acts as a dynamic constraint on the physical structure of the energy momentum transport in the energy momentum tensor $T^{\mu\nu}$ on the right hand side of the Einstein equation $G^{\mu\nu} = (8\pi G / c^4)\, T^{\mu\nu}$.

In this way the Principle of Equivalence dynamically enforced the requirement that the physical surface surrounding a collapsing radiating plasma must obey $1 / (1+ z_S) > 0$ where z is the surface red shift. We found that when this condition is applied to the first integral of the time-time component of the Einstein Equation, in the interior co-moving metric of a collapsing radiating plasma, it leads to the *"no trapped surface condition"* $2GM(r_S , t) / R(r_S , t)c^2 < 1$.

We then argued that Principle of Equivalence enforces this *"no trapped surface condition"* by constraining the physics of the general relativistic radiation transfer process in a manner which requires it to establish and maintain an Eddington limited secular equilibrium on the dynamics of the collapsing radiating surface so as to keep it outside of its Schwarzschild radius during the collapse process and thus prevent the physical surface of the collapsing object from ever forming a trapped surface. We found that the dynamic radiation transfer condition required for the existence of the high redshift Eddington limited stationary equilibrium, associated with the gravitational collapse of a compact object made up of a radiating plasma satisfying the Principle of Equivalence, is that it must contain an intrinsic equipartition magnetic field energy density which is less than or equal to its gravitational field energy density.

---------------------------------------------------------------- Footnotes----------------------------------------------------------

(1)For an example of this, note that the physical speed ratio (V / c) of a particle falling along a radial geodesic in the Schwarzschild metric is given by $(V / c)^2 = \{1 - (1 - 2GM(r) / Rc^2) (mc^2 / E)\}^2$ where m is the particle mass and E is the conserved energy. Then using this result it follows from equation (2-k) that the line element $ds^2$ along this geodesic is $ds^2 = (c\,dt)^2 \{(g_{oo})(1 - V^2/ c^2)\} = (c\,dt)^2\{(1 - 2GM(r) / Rc^2) / (mc^2 / E)\}\}^2 > 0$.. Hence we see that as $2GM(r) / Rc^2$ ----> 1 then $ds^2$ ----> 0. Also for non-zero values of M(r) we find that in this limit the acceleration scalar $a = (a^\mu a_\mu)^{1/2}$ becomes singular which is not allowed in regular regions of spacetime. For a detailed discussion for other coordinates besides the standard Schwarzschild coordinates. see reference [1]

(2)Another possibility is a stellar burst/collapse scenario (where neutrinos are dominantly radiated instead of photons). Then neutrino luminosities $L(\infty) = 10^{57}$ erg/s are possible for a 10 solar mass collapsing object implying a more a rapid collapse scenario . However the nucleons inside the rapidly collapsing surface continue to cool by radiating neutrinos. They eventually cool into a relativistic gas in equilibrium which after a period of time becomes opaque to photons. Finally photon emission dominates neutrino emission and the collapse becomes photon Eddington limited as indicated above.

(3) It has been recently suggested [10] that if the back reaction associated with strong gravity can cause a gravitationally induced Bose-Einstein phase transition of the collapsing matter to occur in the exotic form $p \sim -\rho c^2$ , it may be possible to create static Ultra Compact Objects (UCO) of arbitrary high mass. However this will not occur for sufficiently massive compact plasma objects whose continuous collapse will cause them to heat up and radiate energy, since they will eventually become MECO emitting radiation at the Eddington limit.

(4) It is interesting to note that such strong intrinsic equipartition magnetic fields would tend to suppress type I X-ray bursts from occurring on the highly red shifted surface of the MECO, consistent with what is seen in the case of GBHC.

(5) In contradistinction to that of a rapidly rotating charged Kerr-Newman scenario, the MECO is both charge neutral and assumed to be a slow rotator. In this context the MECO intrinsic magnetic moment will mostly be generated by the nuclear and plasma physics processes going on inside of it

**APPENDIX A: THE EDDINGTON LIMITED COLLAPSE SCENARIO FOR MAGNETOSPHERIC ETERNALLY COLLAPSING OBJECTS (MECO)**

Let us assume an energy momentum tensor which involves matter, pressure, and radiation

$$T_\mu{}^\nu = (\rho + P / c^2)\, u_\mu u^\nu - P\, \delta_\mu{}^\nu + E_\mu{}^\nu \qquad \text{(A1)}$$

where $u^\mu = (u^0\ , 0)$ *in the interior where co-moving coordinates are used* and

$E_\mu{}^\nu = q\, k_\mu\, k^\nu$ is the radiation part of EM tensor in geometric optics limit $k_\mu\, k^\mu = 0$

*Interior co-moving metric*

$$ds^2 = A\,(r\,,t)^2\,(c^2 dt^2) - B(r\,,t)^2\, dR^2 - R(r\,,t)^2 [d\theta^2 + \sin\theta^2 d\phi] \qquad \text{(A2)}$$

*Exterior Radiating metric*

$$ds^2 = a(r\,,u)^2\,(c^2 du^2) + 2\, b(r\,,u)\, c\, du\, dR - R(r\,,u)^2 [d\theta^2 + \sin\theta^2 d\phi^2] \qquad \text{(A3)}$$

where $du = dt - R / c$ is the retarded observer time

Now in the interior co-moving coordinates

$$L = 4\pi\, R^2\, q\, c \qquad \text{luminosity of radiation} \qquad \text{(A4)}$$

$$q = E_\mu{}^\nu\, u_\nu\, u^\mu = E_0{}^0\, c^2 \qquad \text{energy density of radiation} \qquad \text{(A5)}$$

and

$$\Gamma = (dR / dl) \qquad U = (dR / d\tau) \qquad \text{(A6)}$$

$$\Gamma^2 = [1 - R_{Schw} / R + (U / c)^2] \qquad \text{(A7)}$$

From the boundary condition on the time like collapsing surface $r_S = R_S$ (r ,u) separating the interior metric from the exterior metric , and the Principle of Equivalence (POE), we find that the proper time of the collapsing surface is given by

$$d\tau_S = du / (1 + z_S) = du \; [\Gamma_S + U_S / c ] > 0 \qquad \text{(A8)}$$

To analyze the physical gravitational collapse process one proceeds to solve the Einstein equations for the interior co-moving metric (A2), with boundary conditions which connect the interior metric (A2) to the exterior metric (A3) at the collapsing surface S. When this is done, as described by the Einstein equations in the co-moving metric [6,7,8,9], we find that among the various equations associated with the collapse process there are three proper time differential equations which control the physical properties of a compact collapsing and radiating physical surface S. When evaluated on the physical surface S these equations are given by

$$dU_S / d\tau = [\Gamma / (\rho + P/c^2)]_S (- \partial P/ \partial R)_S - G\{[M + 4\pi R^3 (P + q) / c^2] / R^2\}_S \qquad \text{(A9-a)}$$

$$dM_S / d\tau = - [4\pi R^2 P c (U/c)]_S - [L (U/c + \Gamma)]_S \qquad \text{(A9-b)}$$

$$d\Gamma_S / d\tau = G /c^4 (L/ R)_S + [(U/c^2) \{\Gamma / (\rho + P /c^2)\} (- \partial P / \partial R)]_S \qquad \text{(A9-c)}$$

Now we have seen that in all frames of reference the Principle of Equivalence requires that $1 / (1 + z_S) > 0$ must hold in the context of the time like gravitational collapse of physical matter. In other words the Principle of Equivalence implies a surface of infinite red-shift cannot be dynamically formed by the time like collapse of physical matter.

Since the Principle of Equivalence dynamic condition that $1 / (1 + z_S) > 0$ holds for the time like motion of physical matter in all frames of reference, it is also true in the context of the co-moving frame of reference for the equations of motion of the gravitational collapsing surface defined by equations (A9). Hence the Principle of Equivalence implies that equations (A9) must be dynamically constrained to obey

$$1 / (1 + z_S) = \{\Gamma_S + U_S / c\}$$

$$= \{[1 - R_{SCHW} / R + (U / c)^2]^{1/2} + U_S / c\} > 0 \qquad \text{(A9-d)}$$

Solving this equation for the quantity $(R_{SCHW} / R)$ we find this requires that

$$(R_{SCHW} / R_S) < 1 \qquad \text{(A9-e)}$$

must hold in the equation of motion for physical matter.

It follows that the Principle of Equivalence dynamically requires, in equations of motion for physical matter given by (A9), that the condition $R_{Schw} / R_S < 1$ must be maintained by the physical forces which act during the collapse process. Hence we conclude that the Principle of Equivalence requires that the physical processes involved in the gravitational collapse of a physical plasma must allow it to heat up and radiate so as to allow a high red shift Eddington limited secular equilibrium to form in a manner which allows $R_{Schw} / R_S < 1$ to be maintained. This implies for a physical plasma undergoing a spherical gravitational collapse without an surrounding accretion disk, a Magnetospheric Eternally Collapsing Object (MECO) can dynamically form in the context of a high red shift Eddington limited secular equilibrium(1), when both $D_\tau U_S = 0$ and $U_S(\tau_{Edd}) = 0$ occur for the time like motion of the surface layer. The high red shift Eddington limited secular equilibrium occurs at proper time $\tau = \tau_{Edd}$ with invariant surface radius $R_S(\tau_{Edd}) = R_S(Edd) > R_{Schw}(\tau_{Edd})$ whose large but finite value of the surface red shift is given by $[(1 + z_S(\tau_{Edd})]$ .

Then for high red shift Eddington limited MECO equation (A9-a) with $U_S/c) = 0$ becomes

$$dU_S / d\tau = [\{\Gamma_S / (\rho + P/c^2)_S\} (- \partial P/ \partial R)_S - G\{[M + 4\pi R^3(P + q) /c^2]\}_S / R_S^2 = 0 \qquad \text{(A9-a')}$$

Equation (A9-a') when integrated over a closed surface for $\tau \geq \tau_{Edd}$ can be solved for the net outward flow of Eddington limited luminosity thru the surface. Since the surface $R_S(\tau_{Edd})$ obeys $R_{Schw}(\tau_{Edd}) < R_S(\tau_{Edd}) < R_{PHOTON}(\tau_{Edd})$, then $\Gamma_S(\tau_{Edd}) > 0$ and the photon escape cone factor $\sim 27 [R_{Schw}(\tau_{Edd}) /2 R_S(\tau_{Edd})]^2 [1 - R_{Schw}(\tau_{Edd}) /R_S(\tau_{Edd})]] \sim (27/4) / (1 + z_{Edd} )^2]$ must be taken into account in doing the calculation.

When this is done one finds that the out flowing Eddington luminosity emitted from the surface is (see Appendix B) given by

$$L_{Edd}(outflow)_S = \{(4\pi G M_S(\tau) c /\kappa)(1 + z_{Edd} )\} \{(1 + z_{Edd} )^2 / (27/4)\} \qquad \text{(A9-a'')}$$

However the luminosity $(L)_S$ which appears in equations (A9-b,c) is the net luminosity which escapes thru the photon radius which is given by

$$(L)_S = L_{Edd}(escape)_S$$

$$= L_{Edd}(outflow)_S - L_{Edd}(fall\ back)_S$$

$$\sim (L_{Edd})_S - (L_{Edd})_S [1 - (27/4) / (1 + z_{Edd})^2]$$

$$\sim (L_{Edd})_S [(27/4) / (1 + z_{Edd})^2]$$

$$\sim \{(4\pi G M_S(\tau) c / \kappa)(1 + z_{Edd})\} \qquad \text{(A9-a''')}$$

In this context from (A9-b) we have that

$$dM_S / d\tau = - [L_{Edd}(escape)_S / (1 + z_{Edd}) c^2]$$

$$= - [(4\pi G M_S(\tau) c / \kappa) / c^2] = (c^2 / 2G) U_{SCHW} \qquad \text{(A9-b')}$$

which can be integrated for $\tau \geq \tau_{Edd}$ to give

$$M_S(\tau) = M_S(\tau_{Edd}) \exp \{- [(4\pi G / c \kappa (\tau - \tau_{Edd}])\} \qquad \text{(A9-b'')}$$

Finally equation (A9-c) becomes

$$d\Gamma_S / d\tau = G / c^4 (L_{Edd})_S / R_S(\tau_{Edd})$$

$$= (4\pi G / c \kappa) \{[R_{Schw}(\tau) / R_S(\tau_{Edd})] / 2\Gamma_S\} \qquad \text{(A9-c')}$$

whose solution

$$\Gamma_S(\tau) = 1 / (1 + z_S(\tau)) = [(1 - R_{Schw}(\tau) / R_S(\tau_{Edd})]^{1/2} > 0 \qquad \text{(A9-c'')}$$

is consistent with (A9-d).

Hence from the above we have for the high red shift MECO solutions to the Einstein Equations that $R_S(\tau) = R_S(Edd)$, $U(\tau) = U(\tau_{Edd}) = 0$ and

$$R_{Schw}(\tau) = [R_{Schw}(\tau_{Edd})] \exp \{- [(4\pi G / c\, \kappa)(\tau - \tau_{Edd}])\} \qquad \text{(A10-a)}$$

$$< R_S(\tau) = R_S(Edd)$$

$$U_{Schw}(\tau)/ c = [U_{Schw}(\tau_{Edd}) / c] \exp\{- [(4\pi G / c\, \kappa (\tau - \tau_{Edd}])\} \qquad \text{(A10-b)}$$

$$< U(\tau)/ c = U(\tau_{Edd}) / c = 0$$

where

$$[U_{Schw}(\tau_{Edd}) / c] = - \{[ R_{Schw}(\tau_{Edd})] \;/\; [c^2 \kappa /4\pi G]\} \qquad \text{(A10-c)}$$

and

$$\Gamma_S(\tau) = [ (1 - R_{Schw}(\tau) / R_S(\tau_{Edd}) ]^{1/2} = 1 / [1 + z_{Edd}(\tau)] \qquad \text{(A10-d)}$$

$$= [(1 - R_{Schw}(\tau_{Edd})) / R_S(\tau_{Edd}) \exp \{- [(4\pi G / c\, \kappa)(\tau - \tau_{Edd})\}]^{1/2}$$

with

$$[1 + z_{Edd}(\tau)] = [(1 - R_{Schw}(\tau_{Edd})) / R_S(\tau_{Edd}) \exp \{- [(4\pi G / c \kappa)(\tau - \tau_{Edd})\}]^{-1/2}$$

$$\sim [(1 - R_{Schw}(\tau_{Edd})) / R_S(\tau_{Edd}) \exp \{-\alpha(t - t_{Edd})\}]^{-1/2} \qquad \text{(A10-e)}$$

$$\geq [1 + z_{Edd}(\tau_{Edd})] \sim \{(B_{EQUIP}) / (B(\infty)_{EQUIP})\} \sim 10^8$$

Hence a distant observer sees an extremely long e-folding time for the increase of the MECO red shift as

$$\alpha \sim \{1 / [(c \kappa / 4\pi G) (1 + z_S(t_{Edd})]\} \qquad \text{(A10-f)}$$

$$\sim \{1 / [4.5 \times 10^{16} \text{ yrs}]\} = \{1 / [3x10^6 (t_{Hubble})]\}$$

Hence for proper time $\tau \geq \tau_{Edd}$ the MECO mass decays and red shift increases with an extremely long e-folding time of $4.5 \times 10^8 (1 + z_S)$ years $\sim 3x10^6 (t_{Hubble})]\}$.

Finally if $\Delta R_S(\tau) = [R_S(Edd) - R_{Schw}(\tau_{Edd})]$ is the proper radial difference between the physical surface of the MECO and its collapsing Schwarzschild radius given by we have from (A7) and (A10-e) that

$$\Delta R_S(\tau) / R_{Schw}(\tau) = \left[1 / [1 + z_{Edd}(\tau)]^2\right] / \left[1 - 1 / [1 + z_{Edd}(\tau)]^2\right] \qquad \text{(A11)}$$

$$\sim \left[1 / [1 + z_{Edd}(\tau_{Edd})]^2\right] / \left[1 - 1 / [1 + z_{Edd}(\tau_{Edd})]^2\right]$$

$$\sim \left[1 / [1 + z_{Edd}(\tau_{Edd})]^2\right]$$

$$\sim 10^{-16}$$

**APPENDIX B : CALCULATION OF THE GENERAL RELATIVISTIC EDDINGTON LIMIT FORMULAS**

From the equation (A9-c) in the Eddington limit at a surface S

$$\Gamma_{Edd}\,(-dP/\,dR)_{Edd} = [\,(\rho + P/c^2)_S\,]\ [G\,M_S\,/\,R_S{}^2] \qquad \text{(B1-a)}$$

When integrated over a closed surface for $\tau \geq \tau_{Edd}$ equation (B1-a) can be solved for the net outward flow of Eddington limited luminosity thru the surface. Since the surface of invariant radius $R_S(\tau_{Edd})$ obeys $R_{Schw}(\tau_{Edd}) < R_S(\tau_{Edd}) < R_{PHOTON}(\tau_{Edd})$, then $\Gamma_S(\tau_{Edd}) > 0$ and the photon escape cone factor $\sim 27\,[R_{Schw}(\tau_{Edd})\,/2\,R_S(\tau_{Edd})]^2\ [1 - R_{Schw}(\tau_{Edd})\,/R_S(\tau_{Edd})]] \sim$

$(27/4)\,/\,(1 + z_{Edd}\,)^2]$

must be taken into account in doing the calculation. Then integrating this equation over a plasma shell of radius $R_S$

$$\int 4\pi\, R_S{}^2 dR_S\, \Gamma_{Edd}(-dP/\,dR)_{Edd} = \int [4\pi\, G\, M_S]\,[\,(\rho + P/c^2)_S\,]\, dR \qquad \text{(B1-b)}$$

$$\Gamma_{Edd} \left\{ \int 4\pi\, R_S{}^2\, dP_{Edd} \right\} = [4\pi\, G\, M_S] \left\{ \int [\,(\rho + P/c^2)_S\,]\, dR \right\} \qquad \text{(B1-c)}$$

where $\Gamma_{Edd} = 1\,/\,(1 + z_{Edd})$ (B2-a)

and if $N_S$ are the number of protons/cm$^2$ in the plasma shell it follows that

$$\left\{\int [ (\rho + P /c^2)_S ] \, dR \right\} = \{m_P N_S \} \qquad \text{(B2-b)}$$

$$\left\{\int 4\pi \, R_S{}^2 \, dP_{Edd} \right\} = \{\sigma_T \, (L_{Edd}(escape)_S / c)_S \, N_S \} \qquad \text{(B2-c)}$$

Here $L_{Edd}(escape)_S$ is the net luminosity which escapes thru the photon radius given by

$$L_{Edd}(escape)_S = L_{Edd}(outflow)_S - L_{Edd}(fall\ back)_S$$

$$\sim L_{Edd}(outflow)_S - L_{Edd}(outflow)_S \, [1 - (27/4) / (1 + z_{Edd})^2]$$

$$\sim L_{Edd}(outflow)_S \, [(27/4) / (1 + z_{Edd})^2] \qquad \text{(B2-c)}$$

Solving equations (B2-a,b,c) we have locally that

$$L_{Edd}\,(outflow)_S = \{(4\pi \, G \, M_S(\tau) \, c /\kappa)\} \; \{(1 + z_{Edd})^3 / (27/4)\} \qquad \text{(B3-a)}$$

and hence

$$L_{Edd}(escape)_S = \{(4\pi \, G \, M_S(\tau) \, c /\kappa)\} \; \{(1 + z_{Edd})\} \qquad \text{(B3-b)}$$

which implies that red shift at which the Eddington limit is established is given by

$$(1 + z_{Edd}) = \{L_{Edd}\,(outflow)_S / [(4/27)(4\pi \, G \, M_S \, c) / \kappa ]\}^{1/3} \qquad \text{(B3-c)}$$

while the net Eddington luminosity seen at infinity is given by [11, 12]

$$(L_{Edd})_{\infty} = \{L_{Edd}(escape)\}_S \,/\, (1 + z_{Edd})^2$$

$$= [(4\pi G M_S) / \kappa ] \,/\, (1 + z_{Edd}) \qquad \text{(B3-d)}$$

REFERENCES

[1] Mitra , A., Foundations of Physics Letters (2000) , 13, 543, (astro-ph/9910408); and Foundations of Physics Letters (2002 in press) , (astro-ph/0207056)

[2] Robertson, S., Leiter, D., ApJ, (2002), 565, 447 (astro-ph/0102381), and (2002 submitted to ApJ), (astro-ph/0208333)

[3] Einstein, A., (1939), Annals of Mathematics, 40, 922

[4] Landau, L.D., Lifshitz, E. M., (1975), *Classical Theory of Fields*, $4^{th}$ Ed, Pergamon Press pg 248-252

[5] Wheeler, A., Ciufolini, I., “Gravitation and Inertia” , (1995), Princeton University Press Princeton, New Jersey, pp. 69-71.

[6] Hernandez Jr. , W.C., Misner, C.W., ApJ. (1966),` 143, 452

[7] Lindquist, R. W., Schwartz, R. A. , Misner, C. W. , (1965), Phys. Rev. , 137B, 1364.

[8] Misner, . C. W., Phys. Rev. , (1965), 137B, 1360

[9] Lindquist, R. W., Annals of Physics, (1966), 37, 487

[10] Mazur, P.O., Mottola , E. (2001), gr-qc/0109035.

[11] Pelletier, G., & Markowith, A., (1998), ApJ., 502, 598.

[12] Shapiro S., Teukolsky S., (1983) “Black Holes, White Dwarfs, and Neutron Stars”, John Wiley & Sons Inc. New York, see footnote on pg 396.

[13] Mitra, A., (1998) astro-ph/9811402